\documentclass[preprintnumbers,prd,onecolumn,floatfix,superscriptaddress, nofootinbib]{revtex4-2}

\usepackage{setspace}
\usepackage{graphicx}
\usepackage{subfig}
\usepackage{epsfig}
\usepackage{bm}
\usepackage{amssymb}
\usepackage{float}
\usepackage{amsmath}
\usepackage{dcolumn}
\usepackage[colorlinks]{hyperref}
\usepackage[usenames,dvipsnames]{color}
\usepackage{enumitem}

\hypersetup{ breaklinks=true, pdfstartview={FitH}, colorlinks=true, linkcolor=blue, citecolor=red, filecolor=magenta, urlcolor=blue, anchorcolor=green, linktocpage=true }

\def\doi{http://doi.org}

\newcommand{\be}{\begin{equation}}
\newcommand{\ee}{\end{equation}}
\newcommand{\beano}{\begin{eqnarray*}}
\newcommand{\eeano}{\end{eqnarray*}}
\newcommand{\ba}{\begin{eqnarray}}
\newcommand{\ea}{\end{eqnarray}}

\begin{document}

\title{ The constrained cosmological model in Lyra geometry}

\author{J. K. Singh}
\email{jksingh@nsut.ac.in}
\affiliation{Department of Mathematics, Netaji Subhas University of Technology, New Delhi-110078, India}
\author{Shaily}
\email{shailytyagi.iitkgp@gmail.com}  
\affiliation{Department of Mathematics, Netaji Subhas University of Technology, New Delhi-110078, India}
\author{Shri Ram}
\email{srmathitbhu@rediffmail.com}
\affiliation{Department of Mathematical Sciences, Indian Institute of Technology (BHU), Varanasi 221 005, India}
\author{Joao R. L. Santos}
\email{joaorafael@df.ufcg.edu.br}
\affiliation{UFCG-Universidade Federal de Campina Grande-Unidade Academica de Fisica, 58429-900 Campina Grande, PB, Brazil}
\author{Jéferson A. S. Fortunato}
\email{jeferson.fortunato@edu.ufes.br}\affiliation{PPGCosmo, CCE, Universidade Federal do Espírito Santo (UFES), 29.075-910, Vitória, ES, Brazil}

\begin{abstract}
\begin{singlespace}

In this article, we study a flat homogeneous FLRW model in Lyra geometry which is described by a time-dependent displacement vector. We consider an appropriate parametrization of the energy density of scalar field $ \rho_\phi $ in terms of the cosmic scale factor. The result shows two transitions from deceleration to acceleration. Furthermore, we constrain the model parameter $ \alpha $ and the displacement field vector $ \beta $ using the recent supernovae data, Hubble data set of 77 points and their joint data which predicts the accelerated expanding phase of the universe in late times. The effective equation of state parameter $ \omega_{eff} $ speculate $ \Lambda $CDM in late times. Finally, we use the statefinder diagnostic to differentiate our model from the various dark energy models.

\end{singlespace}
\end{abstract}

\maketitle

\section{Introduction} 

\qquad From various observations, the accelerating behavior of the Universe is confirmed. However, figuring out the reason for acceleration is still a big challenge in cosmology. In standard cosmology, dark energy models are supported as a resolution to this fundamental question. Here, the cosmological constant $ \Lambda $, which can be associated with the vacuum quantum energy, is most prominent to describe the dark energy \cite{Weinberg:1988cp,Copeland:2006wr}. Another reassuring way to understand the recent probes for cosmic expansion is Einstein theory of general relativity \cite{Buchdahl:1970ynr,Kerner:1982yg}. The Einstein’s GR models seize up at large cosmic scales and a more generic action characterizes the gravitational field. To generalize the Einstein-Hilbert action of GR, there are various models where standard action is replaced by the generic function $ f(R) $, where $ R $ is the Ricci scalar \cite{Kleinert:2000rt, Carroll:2003wy, Capozziello:2006dj,Amendola:2006kh,Nojiri:2003ft,Faraoni:2006hx,Zhang:2007ne,Amendola:2007nt}. Using these $ f(R) $ models, the early inflation of the universe and the concept of dark energy are discussed. In the extension of $ f(R) $ gravity theory, an explicit coupling of matter Lagrangian density $ L_m $ is included with $ f(R) $ \cite{Tsujikawa:2007xu,Liu:2017xef,Capozziello:2007eu,Starobinsky:2007hu,Nojiri:2007as,Nojiri:2007cq,Cognola:2007zu,Santos:2007bs,Capozziello:2008qc, Bertolami:2007gv,Harko:2008qz}. For a proper description of the early universe, homogeneous and isotropic cosmological models play a significant role. 

It is noteworthy that among all of the alternative theories of gravitation, Lyra geometry is one of the important theory, which is proposed by Lyra \cite{Lyr}. Lyra geometry comes with a modification of Riemannian geometry and may be considered as a modification of Weyl’s geometry. In Weyl's geometry, the change in the length of vectors under parallel transport is non-zero and depends on a new vector quantity $ \zeta_i $, which is used for electromagnetic potential. In 1951, Lyra proposed a different way to maintain the vector length integrability by adopting a gauge function $ \beta(t) $ as an intrinsic part of the manifold’s geometric structure \cite{Cuzinatto:2021ttc,Penzias:1965wn,Bha, Bee, Hal, Red,Singh:1991zz, Nuo,Singh:1992wd,Sen1, Sen:1971fac, Singh:1997zz}. Further, Sen \cite{Sen} formulate a static cosmological model where he obtained the EFE through the variational principle. Singh and Shri Ram \cite{Singh:2009zzb} discussed the spatially homogeneous Bianchi type-I metric in the normal gauge for Lyra’s geometry.

In the literature, a reconstructing technique or model-independent technique has been employed. Nowadays, this model-independent technique is of great interest to configure of some DE candidates. Initially, this approach is discussed by Starobinsky \cite{Starobinsky:2007hu}. Many authors proposed this parametrization scheme on density, pressure, deceleration parameter, Hubble parameter, and scale factor. This method has two categories: (1) non-parametric and (2) parametric. The non-parametric method has certain restrictions because it could not be able to explain all the incidents of the present Universe while the parametric method includes a specific parametrization of the distinct types of cosmological parameters which are used to explore the cosmological models \cite{Singh:2019fpr}.

The present work is organized as follows: In Sec. II, we present the model field equations using Lyra's geometry frame transformation for flat FLRW space-time metric. We parametrize the energy density for the scalar field and obtain the analytical solutions of the Hubble parameter and other cosmological parameters. In Sec. III, we discuss a brief introduction of Type Ia supernova data and observational Hubble data to derive the constraints on the model parameters. In Sec. IV, we study the evolution of various quantities of the universe. Finally, we discuss and conclude our findings in Sec. V.
\section{Field Equations in Lyra geometry}
\qquad The gravitational action proposed by Sen \cite{Sen} is defined as

\begin{equation} \label{1}
A=\int (\beta^4 R+L_\phi+L_m) \sqrt{-g}  d^{4}x,
\end{equation} 

Using Lyra’s reference frame transformation the field equations obtained as:
\begin{equation}\label{2}
R_{\gamma \delta}-\frac{1}{2} R g _{\gamma \delta}+ \frac{3}{2}\zeta_\gamma \zeta_\delta-\frac{3}{4}g_{\gamma \delta} \zeta_\lambda \zeta^\lambda= \kappa T_{\gamma \delta},
\end{equation}

where $ \kappa=8\pi G $ which is normalized to $ 1 $ for further calculation, $ R_{\gamma \delta} $ is Ricci tensor, $ R $ is scalar curvature, and $ \zeta_\gamma $ is the time-like displacement vector, defined as $ (\beta(t),0,0,0) $.

The flat FLRW space-time metric is given by
\begin{equation}\label{3}
ds^2=-a^{2}(t)(dx^2+dy^2+dz^2)+dt^2,
\end{equation}
where $ a(t) $ is the scale factor of the universe. From Eq. (\ref{2}), we get

\begin{equation}\label{4}
3 H^2 - \frac{3\beta^2}{4} = \rho_{eff}=\rho_{\phi}+\rho_m,
\end{equation}
\begin{equation}\label{5}
2 \dot{H}+3 H^2+\frac{3\beta^2}{4} = -p_{eff}=p_{\phi}.
\end{equation}
Overhead dot is used for derivative \textit{w.r.t.} time. $ \rho_{eff} $ and $ p_{eff} $ signifies the total amount of energy density and pressure contained in the universe. Also $ \rho_{\phi} $, $ \rho_m $, $ p_{\phi} $ depict the energy density of the scalar field, energy density of matter, and pressure of scalar field respectively. According to the present Universe is filled with two types of fluid, one is corresponding to the scalar field and another is pressureless cold matter. 

The action of the scalar field is defined as:

\begin{equation}\label{6}
L_\phi=\int [\frac{1}{2}\partial_\nu{\phi}\partial^\nu{\phi}-V(\phi)]\sqrt{-g}d^{4}x.
\end{equation}

In the study regarding an isotropy and homogeneity of the model, we require a dominant unit of the scalar field $ \phi $. As dependency of scalar field is on time therefore this may assume as perfect fluid with $ \rho_{\phi} $ and $ {\phi} $. And for FLRW cosmology energy density and pressure for scalar field can be obtained as

\begin{equation}\label{7}
\rho_{\phi}= \frac{\dot{\phi}^2}{2}+V(\phi),
\end{equation}
\begin{equation}\label{8}
p_{\phi}= \frac{\dot{\phi}^2}{2}-V(\phi),
\end{equation}
where $ V(\phi) $ and $ \frac{1}{2}\dot{\phi}^2 $ are potential energy and kinetic energy depending on scalar field $\phi$. Now the conservation equations for matter field and scalar field read as
\begin{equation}\label{9}
\dot{\rho_m}+3H\rho_m=0,
\end{equation}
\begin{equation}\label{10}
\dot{\rho_{\phi}}+3H(\rho_{\phi}+p_{\phi})=0.
\end{equation}

As EoS parameter $\omega_{\phi}=\frac{p_{\phi}}{\rho_{\phi}}$, thus Eq. (\ref{10}) leads
\begin{equation}\label{11}
\dot{\rho_{\phi}}+3H(1+\omega_{\phi})\rho_{\phi}=0,
\end{equation}
or
\begin{equation}\label{12}
\omega_{\phi}= -( 1+a \frac{1}{3\rho_{\phi}}\frac{d\rho_{\phi}}{da}) ,
\end{equation}

Since we have four unknowns in three independent Eqs. (\ref{9}), (\ref{10}), (\ref{12}), thus in order to solve the system of independent equations we need one more constraint equation. Therefore, we consider an appropriate parametrization of $ \rho_{\phi} $ as
\begin{equation}
\rho_{\phi}(a)=e^{-\alpha a} tan^{-1}(a^{-\alpha}).
\end{equation}\label{13}
This kind of approach is also been used by Singh et al. \cite{Singh:2019fpr}. In this expression $ \alpha \in (0,0.2) $ is the model parameter and will be constrained from observational datasets. The value of $ \rho_{\phi} $ can be written in terms of redshift by using the relation  $ a=\frac{a_0}{(1+z)} $, where $ a_0=1 $ is the present value of scale factor, as
\begin{equation}\label{14}
\rho_{\phi}(z)=e^{\frac{-\alpha}{1+z}} tan^{-1}(1+z)^{\alpha},
\end{equation}
and the present value of energy density \textit{i.e.} at z=0,
\begin{equation}\label{15}
\rho_{\phi_{0}}=\frac{\pi}{4}e^{-\alpha}.
\end{equation}    
Eq. (\ref{14}) and (\ref{15}) yield
\begin{equation}\label{16}
\rho_{\phi}(z)= \frac{4}{\pi}\rho_{\phi_{0}} e^{\frac{\alpha z}{1+z}} tan^{-1}(1+z)^{\alpha},
\end{equation}
Also using Eq. (\ref{9}), energy density of matter field $ \rho_{m} $ can be calculated in terms of redshift $ z $ as 
\begin{equation}\label{17}
\rho_m(z)= \rho_{m_{0}} (1+z)^3
\end{equation}

Together Eqs. (\ref{4}), (\ref{16}) and (\ref{17}),  we have
\begin{equation}\label{18}
 3H^2= \frac{3\beta^2}{4}+\rho_{m_{0}}(1+z)^3+ \frac{4}{\pi}\rho_{\phi_{0}} e^{\frac{\alpha z}{1+z}} tan^{-1}(1+z)^{\alpha},
\end{equation}

Now, by taking the gauge function $ \beta^2=3\beta_0 H_0^2 a^{-2\alpha} $ and considering the density parameter $ \Omega=\frac{\rho}{\rho_c} $ which plays the key role to explain the whole content of the Universe. Here $ \rho_c=\frac{3H^2}{8\pi G} $ is the critical density of the Universe. 

Eq. (\ref{18}) can be written in terms of density parameter of scalar field and matter,
\begin{equation}\label{19}
H= \sqrt{\frac{1}{4}\beta_0 (1+z)^{2\alpha}+\frac{4}{3 \pi}\Omega_{\phi_0} e^{\frac{\alpha  z}{z+1}} \tan ^{-1}(z+1)^{\alpha}+\frac{1}{3}{\Omega_{m_0}} (z+1)^3},
\end{equation} 
where $ H_0 $, $ \Omega_{m_0} $ and $ \Omega_{\phi_0}$ are the current values of Hubble parameter, density parameters of matter and scalar field respectively. We consider the Hubble parameter $ H $ as dimensionless quantity by dividing Eq. (\ref{18}) by $ \sqrt{3} H_0 $, which yields Eq. (\ref{19}). 

The expression for deceleration parameter $ q $ given by  
\begin{equation}\label{20}
q=-\frac{a\ddot{a}}{\dot{a}^2}=-1+\frac{1+z}{H}\frac{dH}{dz}.
\end{equation}

Thus, the deceleration parameter $ q $ and EoS parameter for scalar field $ \omega_{\phi} $ can be evaluated as
\begin{equation}\label{21}
q(z) = -\frac{-\frac{1}{2} \alpha  \beta_0  (\frac{1}{z+1})^{-2 \alpha -1}-\frac{4 \alpha  \Omega_{\phi_0}  e^{\alpha  (1-\frac{1}{z+1})} (\frac{1}{z+1})^{-\alpha -1}}{3 \pi  ((\frac{1}{z+1})^{-2 \alpha }+1)}-\frac{4 \alpha  \Omega_{\phi_0}  e^{\alpha  (1-\frac{1}{z+1})} \tan ^{-1}((\frac{1}{z+1})^{-\alpha })}{3 \pi }-\Omega_{m_0} (z+1)^4}{2 (z+1) (\frac{1}{4} \beta_0  (\frac{1}{z+1})^{-2 \alpha }+\frac{4 \Omega_{\phi_0}  e^{\alpha  (1-\frac{1}{z+1})} \tan ^{-1}((\frac{1}{z+1})^{-\alpha })}{3 \pi }+\frac{1}{3} \Omega _{m_0} (z+1)^3)}-1,
\end{equation} 
 
\begin{eqnarray}\label{22}
\omega_{\phi}(z)=-\frac{\pi  e^{\alpha  (\frac{1}{z+1}-1)} (-\frac{4 \alpha  \Omega \phi  e^{\alpha  (1-\frac{1}{z+1})} (\frac{1}{z+1})^{-\alpha -1}}{\pi  ((\frac{1}{z+1})^{-2 \alpha }+1)}-\frac{4 \alpha  \Omega_{\phi_0}  e^{\alpha  (1-\frac{1}{z+1})} \tan ^{-1}((\frac{1}{z+1})^{-\alpha })}{\pi })}{12 \Omega_{\phi_0}  (z+1)^2 \tan ^{-1}((\frac{1}{z+1})^{-\alpha })}\notag  \\  \times\frac{ \sqrt{\frac{1}{4} \beta_0  (\frac{1}{z+1}t)^{-2 \alpha }+\frac{4 \Omega \phi  e^{\alpha  (1-\frac{1}{z+1})} \tan ^{-1}((\frac{1}{z+1})^{-\alpha })}{3 \pi }+\frac{1}{3} \Omega_{m_0} (z+1)^3}}{12 \Omega_{\phi_0}  (z+1)^2 \tan ^{-1}((\frac{1}{z+1})^{-\alpha })}-1.
\end{eqnarray}

In the next section, we constrain the model parameter $ \alpha $ and $ \beta_0 $ using different observational datasets and obtain their best fit value. Thereafter use these values to examine the behaviours of various physical parameters.

\section{Statistical observation for model parameters} \label{sec:floats} 

\qquad Observational cosmology plays a major role to explore the evolution of the Universe. Therefore, in the current section, we are using the expression for the Hubble parameter from Eq. (\ref{19}) to perform statistical analysis with three observational datasets namely the Hubble dataset (77 points) \cite{Shaily:2022enj}, Type Ia Supernovae dataset \cite{Pan-STARRS1:2017jku} and their joint datasets and to acquire the present best fit value of the model parameters $ \alpha $ and $ \beta_0 $. Here, for the likelihoods minimization of $ \chi^2 $ for the goodness of fit of the model, we use the Markov chain Monte Carlo (MCMC) method to investigate the parameter space from the python package emcee \cite{Foreman-Mackey:2012any}, which is extensively used in observational cosmology.

In this process, the present value of Hubble parameter, density parameters of matter, and scalar field are taken as $ 67.4 KmSec^{-1}Mpc^{-1} $, 0.315, 0.685 respectively, which are given in recent Planck data \cite{Planck:2018vyg}.
\begin{figure}
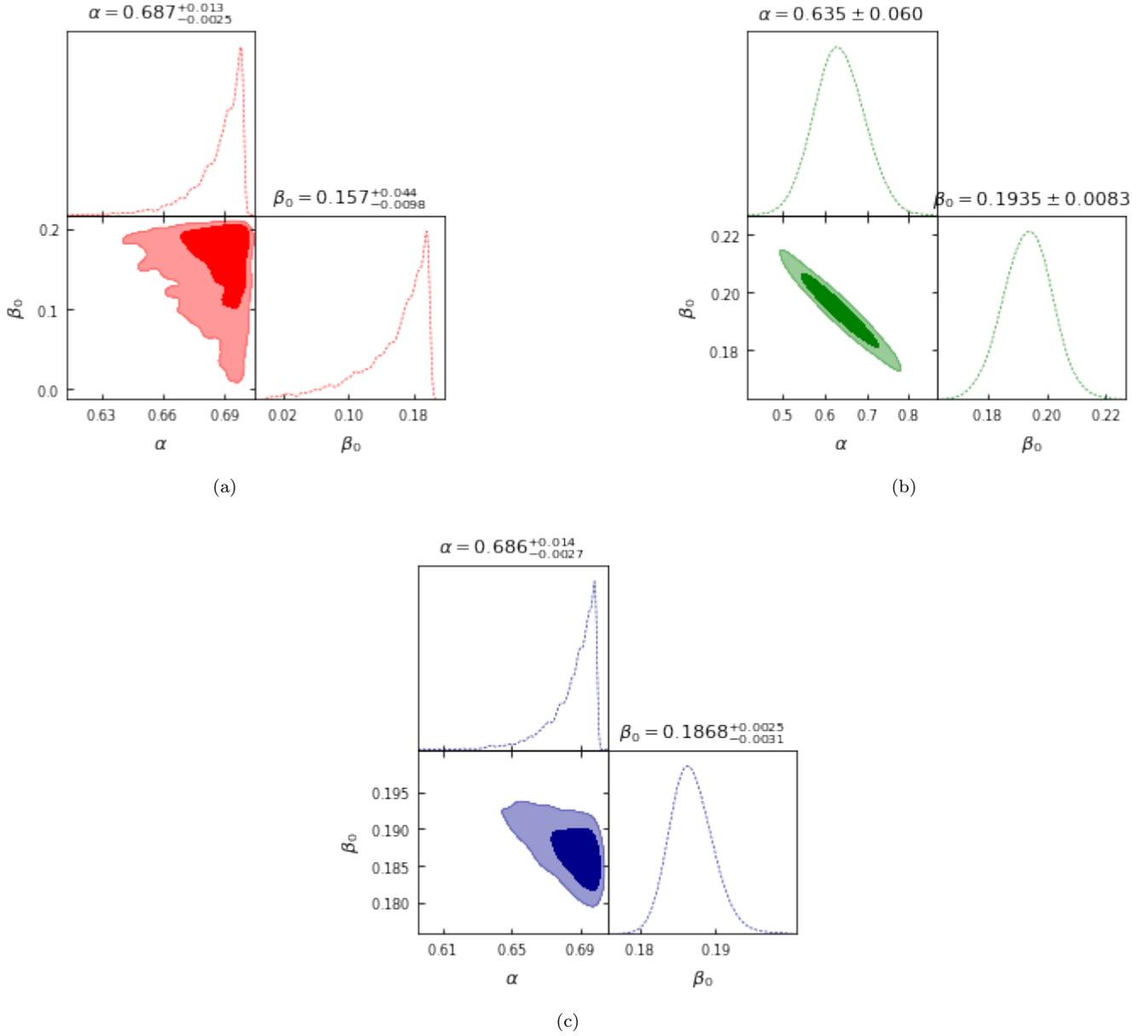
\centering
	\subfloat[]{\label{a}\includegraphics[scale=0.75]{HZ-c1}}\hfill
	\subfloat[]{\label{b}\includegraphics[scale=0.75]{SN-c1}}\par
	\subfloat[]{\label{c}\includegraphics[scale=0.75]{SN+HZ-c1}}
\caption{\scriptsize Figs. (a), (b) and (c) illustrate the constraints on the model for the data $ OHD $, $ SNeIa $, and joint dataset $ OHD+SNeIa $ respectively corresponding to $ 1\sigma $ and $ 2\sigma $ confidence regions.}
\end{figure}
\subsection{Supernovae type Ia Data}
\qquad We use $ SNeIa $ Union $ 2.1 $ compilation dataset and constrained both the model parameters. For flat universe Chi-square function $ \chi_{SN}^{2}(\alpha,\beta_0, H_0) $ can be stated as
\begin{equation}\label{23}
\chi _{SN}^{2}(\alpha,\beta_0, H_0)=\sum\limits_{i=1}^{580}\left[ \frac{\mu_{th}(\alpha,\beta_0,H_0,z_{i})-\mu_{obs}(z_{i})}{\sigma _{\mu(z_{i})}}\right] ^2.
\end{equation}
where $ SN $ is stand for the observational $ SNeIa $ dataset. $ \mu_{th} $ and $ \mu_{obs} $ represents theoretical and observed distance modulus of the model. The standard error in the observed value is denoted by $\sigma_{\mu(z_{i})}$. Also, the theoretical distance modulus  $\mu(z)$ is defined by 
\begin{equation}\label{24}
\mu(z)= m-M = 5Log D_l(z)+\mu_{0},
\end{equation}
where $ m $ and $ M $ are used for the apparent and absolute magnitudes of a standard candle respectively. The luminosity distance $ D_l(z) $ for flat Universe and the nuisance parameter $ \mu_0 $ are given by
\begin{equation}\label{25}
D_l(z)=(1+z) c \int_0^z \frac{1}{H(z^*)}dz^*,
\end{equation}
and
\begin{equation}\label{26}
\mu_0= 5Log\Big(\frac{H_0^{-1}}{1Mpc}\Big)+25,
\end{equation}
respectively. Here, we perform a global fitting to dictate the model parameters using the Markov chain Monte Carlo method and Python implementation of the ensemble sampler for the MCMC method with the EMCEE library, introduced by Foreman-Mackey et al. to show the best-fit value of model parameters written in Table-I.

\subsection{Hubble Dataset}
\qquad Since Hubble data is directly based on differential ages of the galaxies and related to the expansion history of the universe thus this is very useful to understand the dark section \textit{i.e.} dark energy, dark matter, and dark ages of the universe. The value $ H $ can be written in terms of $ z $ as
\begin{equation}\label{27}
H(z) = -(1+z)^{-1} \frac{dz}{dt}.
\end{equation}
The best-fit values of the model parameters are determined by minimizing the Chi-square value 
\begin{equation}\label{28}
\chi _{HD}^{2}(\alpha,\beta_0)=\sum\limits_{i=1}^{77} \frac{[H(\alpha,\beta_0,z_{i})-H_{obs}(z_{i})]^2}{\sigma _{z_i}^2},
\end{equation}
where $ H(\alpha, \beta_0, z_{i}) $ and $ H_{obs}$ represent the theoretical and observed values respectively and $ \sigma_{z_{i}} $ indicates the standard deviation for every $ H(z_i) $. For $ H(z) $ 77 data points, the best fit values of $ \alpha $ and $ \beta_0 $ are estimated in Table-I.

\begin{table}[H]
\caption{Best fit values of model parameters.}
\begin{center}
\label{tabparm}
\begin{tabular}{l c c c r} 
\hline\hline
{Dataset} &       ~~~~~ $ \alpha $  & ~~~~~  $ \beta_0 $ 
\\
\hline      
{$ SNeIa $ }     &  ~~~~~ $ 0.635^{+0.060}_{-0.060} $   &  ~~~~~ $ 0.1935^{+0.0083}_{-0.0083} $   
\\
\\
{$ H(z) $ }     &  ~~~~~ $ 0.687^{+0.013}_{-0.0025 } $   &  ~~~~~ $ 0.157^{+0.044}_{-0.0098} $  & 
\\
\\
{$ H(z) $ + $ SNeIa $ }  &  ~~~~~$ 0.686^{+0.014}_{-0.0027} $  & ~~~~~ $ 0.1868^{+0.0025}_{-0.0031} $  
\\ 
\hline
\end{tabular}    
\end{center}
\end{table}

\textbf{The Gelman-Rubin convergence test \cite{Gelman:1992zz} is a widely used statistical tool in Bayesian inference that allows to access the convergence of the chains. The test is based on the idea that multiple MCMC chains with different starting points should converge to the same posterior distribution if they have been run for long enough, that is, after a number of steps. Basically, this test measures, for each parameter of the considered model, the quantity called potential scale reduction $\hat R$, which is the ratio between the variance $\rm W$ within a chain and the variance $\rm Var(\theta)$ among the chains:
\begin{equation}
    \hat R=\sqrt{\frac{\rm \hat Var(\theta)}{W}}.
\end{equation}
The value of $\hat R$ for a perfectly converged chain should be $1$, so $\hat R \approx 1.1$ corresponds to the maximum allowed value for this parameter for which the convergence of the chain is achieved. }

\begin{figure}[h!]
 \includegraphics[scale=0.50]{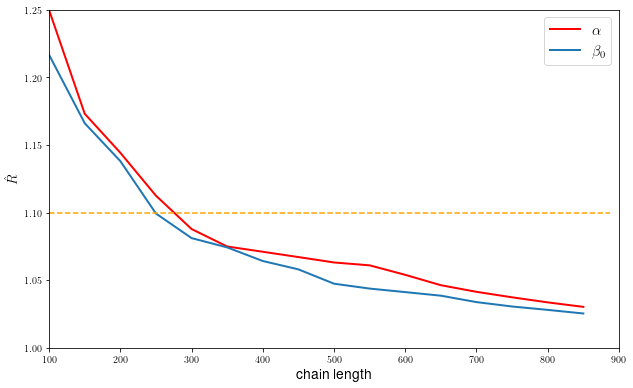}
 \caption{\scriptsize The evolution of the potential scale reduction $\hat R$ with the chain length for $\alpha$ and $\beta$.}
 \label{hatr}
\end{figure}

\textbf{We then apply the Gelman-Rubin convergence test to chains of the fits of $\alpha$ and $\beta_0$ from Hubble data alone. The test yielded $\hat R = 1.03008001$, for $\alpha$ and $\hat R = 1.02498708$ for $\beta_0$, respectively. As $1000$ steps with $100$ steps of burn-in were used to run the MCMC chains, in Fig. \ref{hatr} we verify that convergence, for both parameters, is reached just after approximately $250$ steps when crossing the horizontal line that represents the convergence criterion.}

\subsection{Joint Datasets ($ H(z)+SNeIa $)}
\qquad The $ \chi^2 $ function for joint analysis is given by
\begin{equation}\label{29}
\chi_{HS}^{2}=\chi _{HD}^{2}+\chi _{SN}^{2}.
\end{equation} 

Using this joint statistical analysis, the stronger constraints of the model parameter can be obtained. For joint datasets, the best fit values of $ \alpha $ and $ \beta_0 $, can be seen in Fig. 1(c) and Table-I. Now to compare our model with $ \Lambda CDM $, we use error bar plots for Hubble datasets and Supernovae datasets. The values of model parameters are constrained from observational datasets. In Fig. 3, the error bar plots of the Hubble and Type Ia Supernovae dataset show the best fit plots.
\begin{figure}\centering
	\subfloat[]{\label{a}\includegraphics[scale=0.40]{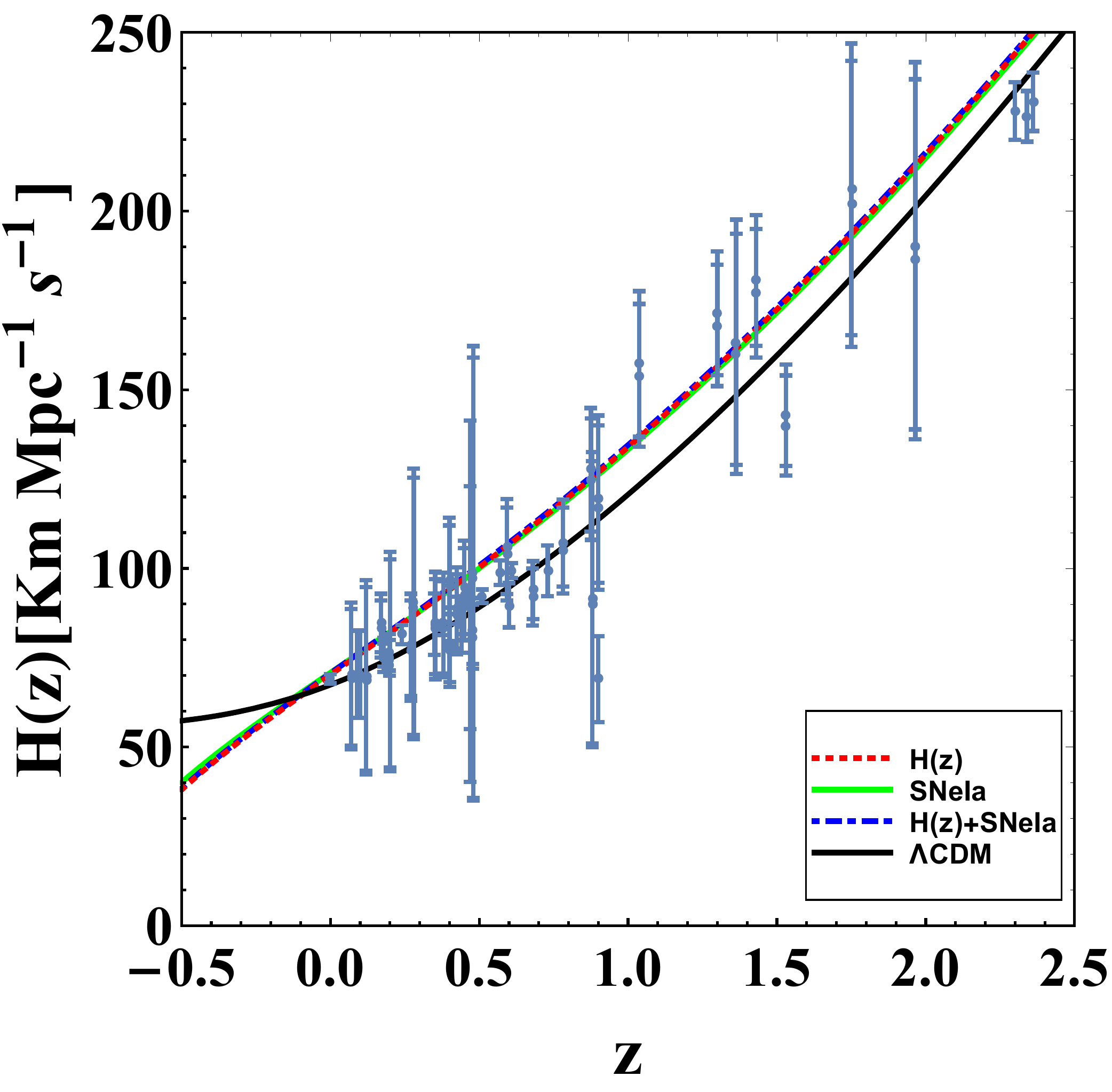}}\hfill
	\subfloat[]{\label{b}\includegraphics[scale=0.40]{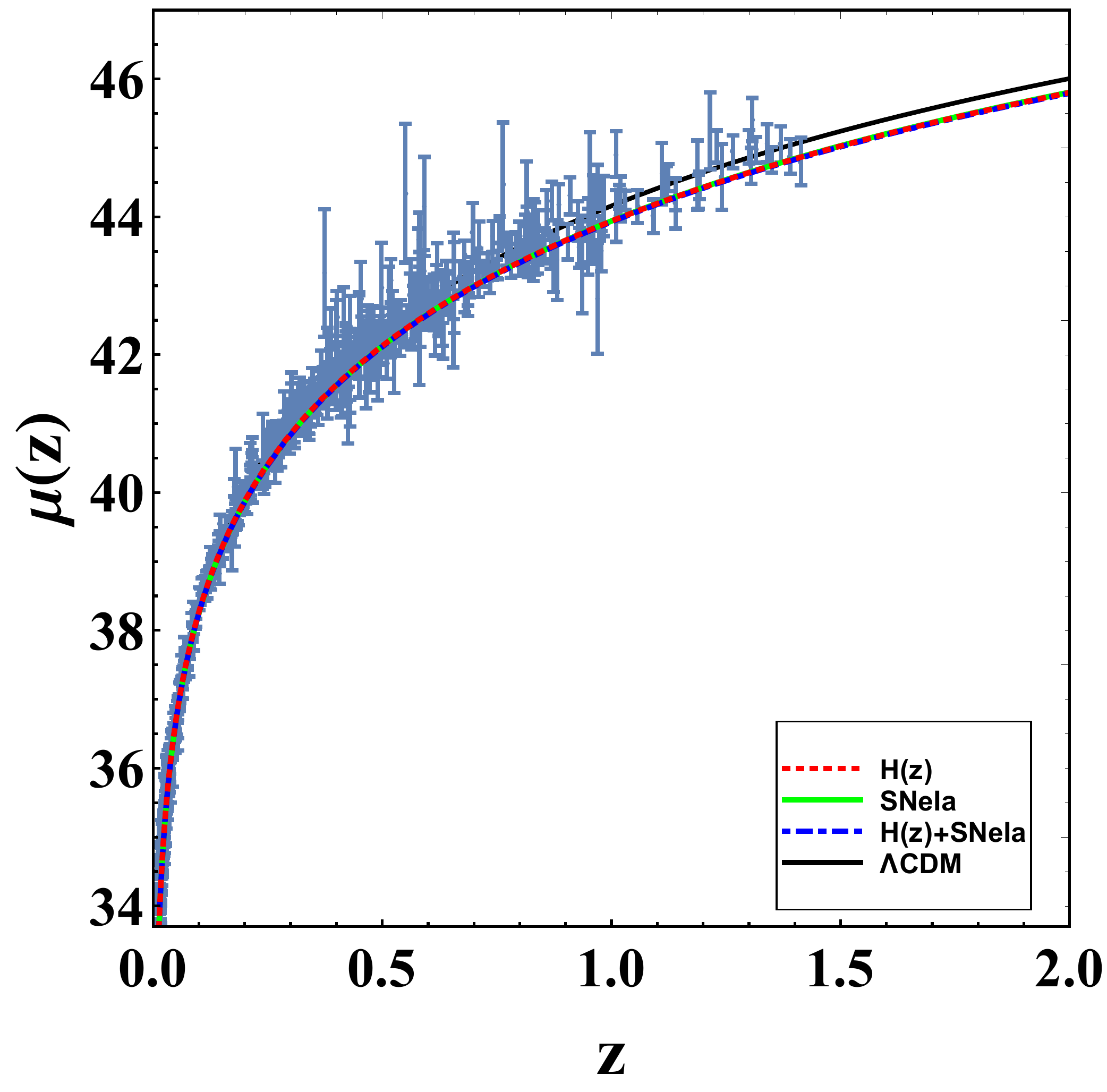}} 
\caption{\scriptsize The error bar plots in respect of $ H(z) $ and Type Ia Supernovae datasets show the similarity between our model and $ \Lambda $CDM.}
\end{figure}

\section{Evolution of the Universe}
 
\qquad In this section, we study the evolution of various cosmological quantities using the best-fit values of the model parameters obtained from different observational datasets. Energy density for the scalar field is decreasing with respect to redshift $ z $. In Fig. 4(a), It is clearly visible that $ \rho_{\phi} $ monotonically decreases and approaches zero in late times, which indicates that amount of dark energy density of the scalar field diminished in the future. In Fig. 4(b), the isotropic pressure of the scalar field is negative for all values of $ z $, which indicates the presence of dark energy up to late times. 
\begin{figure}\centering
	\subfloat[]{\label{a}\includegraphics[scale=0.30]{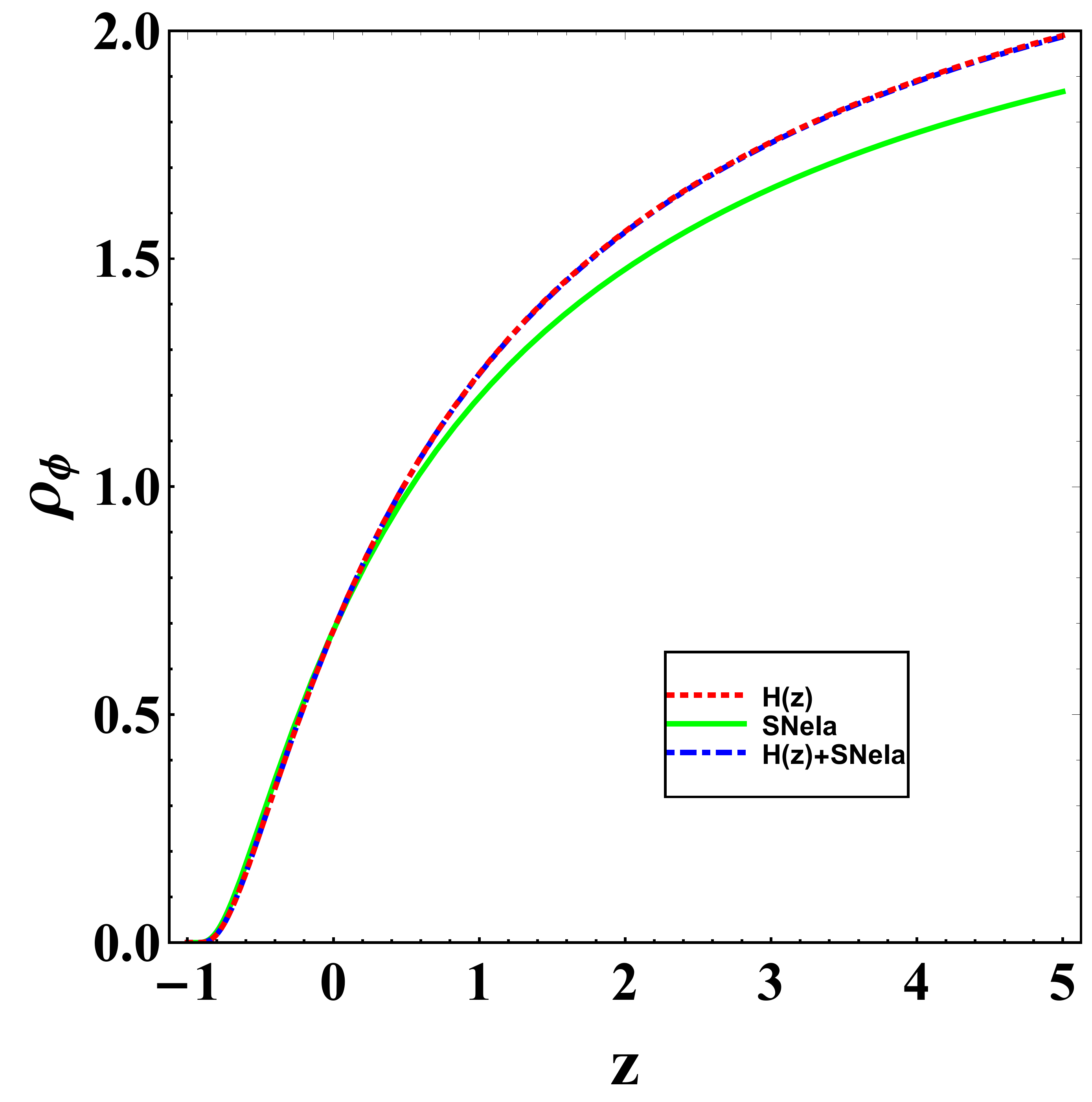}}\hfill
	\subfloat[]{\label{b}\includegraphics[scale=0.32]{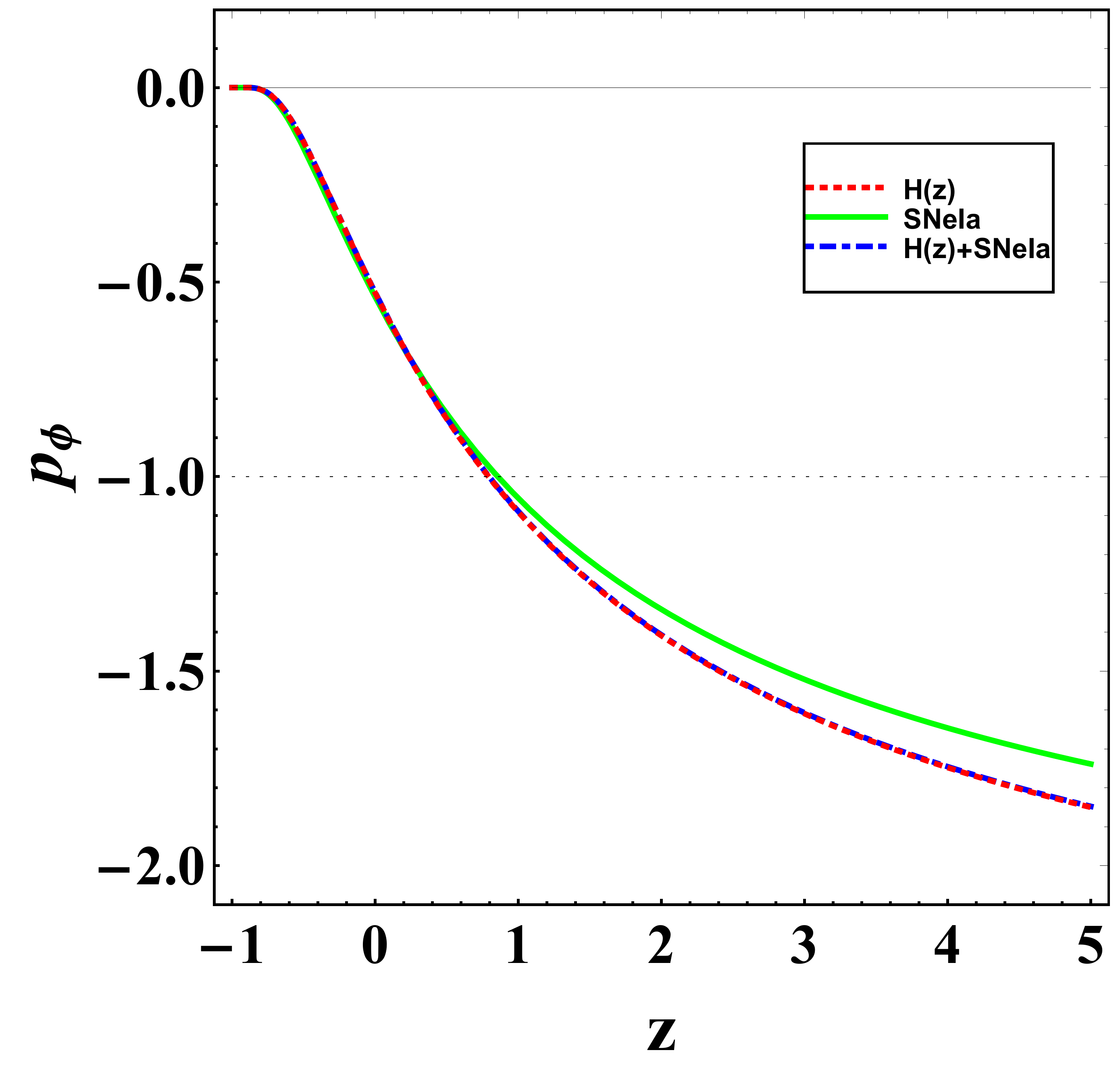}}
	\caption{\scriptsize The graphs of $ \rho_{\phi} $ and $ p_{\phi} $ \textit{w.r.t.} $ z $.}
\end{figure}
Also the plots of EoS parameter $ \omega_{\phi}$ and $ \omega_{eff}=\frac{p_{\phi}}{\rho_{\phi}+\rho_m} $ show the quintessence model in late times. Finally, our model shows the quintessence model in late times as it transits from the quintessence region to a perfect fluid state and again converges to the quintessence region in the future (see Fig. 5).  

\begin{table}[H]
\caption{ The present values of the cosmological parameters.}
\begin{center}
\label{tabparm}
\begin{tabular}{l c c c r} 
\hline\hline
{Dataset} &  ~~~~~~$ H $    &      ~~~~~~~ $ q $ & ~~~~~  $ \omega_{eff} $  
\\
\hline      
{$ SNeIa $ }  &   ~~~~~~ $  70.802  $   &  ~~~~~~~ $ -0.1961 $   &  ~~~~~ $ -0.5283  $  
\\
\\
{$ H(z) $ }    & ~~~~~~  $   70.2202   $    &  ~~~~~~~ $ -0.1604 $   &  ~~~~~ $ -0.5384 $    
\\
\\
{$ H(z) $ + $ SNeIa $ }  &  ~~~~~~$  70.6396   $  &  ~~~~~~~$ -0.1640 $  & ~~~~~ $ -0.5270 $    
\\ 
\hline
\end{tabular}    
\end{center}
\end{table}
\begin{figure}\centering
	\subfloat[]{\label{a}\includegraphics[scale=0.30]{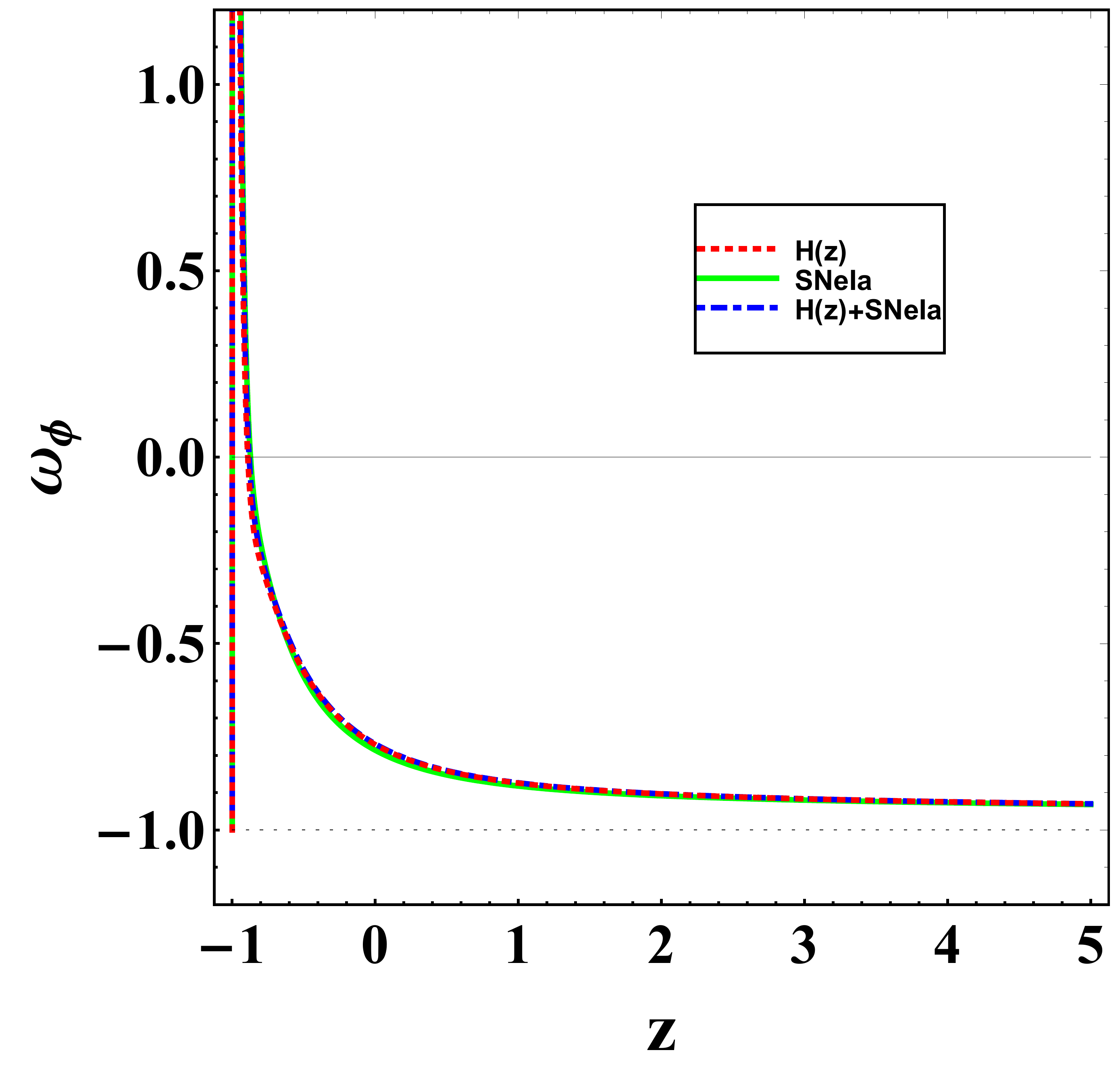}} \hfill
	\subfloat[]{\label{b}\includegraphics[scale=0.30]{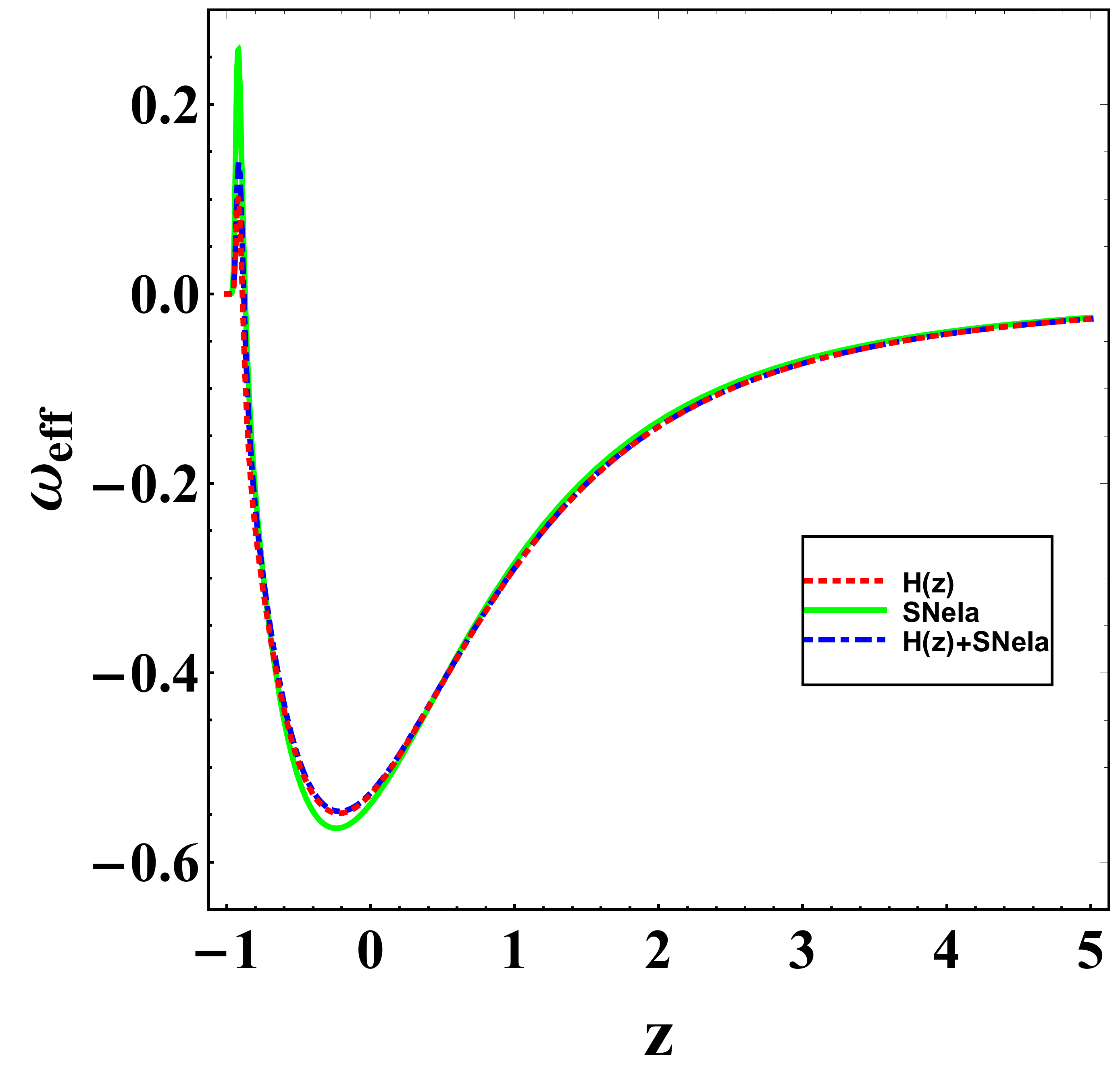}}
\caption{\scriptsize The graphs of $ \omega_{\phi} $ and $ \omega_{eff} $ \textit{\ vs.}  $ z $ for distinct observational data.}
\end{figure}

Fig. 6(a) shows the evolution of the deceleration parameter for different observational datasets. In this model, we observe that the Universe has three transit points during the evolution of each observational data. In early time universe transited from deceleration to acceleration, at present, it shows the acceleration phase. In late time again the universe is having a transition from acceleration to deceleration and afterwords deceleration to acceleration. These transition points are shown in Table II. \textbf{In our model, we observe that the present value of $q$ is not very near to 0.5 and it is relevant to point out that $q \approx - 0.5$ stands for $\Lambda$CDM model and constraints over this parameter using the CMB data. Therefore, new theories of gravity may admit different values for the deceleration parameter, as we can see in the following works \cite{Singh:2018cip, Sahu:2016ccd}. Moreover, the crises over $H_0$ in different regimes of redshift may also open the possibility of different values for $q_0$ and $\omega_0$, as we can see in \cite{Alvarez:2022mlf, Camarena:2019moy}. }

\begin{table}[H]
\caption{ Transition points of deceleration parameter for different datasets.}
\begin{center}
\label{tabparm}
\begin{tabular}{l c c c r} 
\hline\hline
{Dataset} &  ~~~~~~$ z_{tr_1} $    &      ~~~~~~~ $ z_{tr_2} $  & ~~~~~  $ z_{tr_3} $   
\\
\hline      
{$ SNeIa $ }  &   ~~~~~~ $  0.7578  $   &  ~~~~~~~ $ -0.5815 $   &  ~~~~~ $ -0.8824  $    
\\
\\
{$ H(z) $ }    & ~~~~~~  $   0.7093   $    &  ~~~~~~~ $ -0.4942 $   &  ~~~~~ $ -0.8824 $     
\\
\\
{$ H(z) $ + $ SNeIa $ }  &  ~~~~~~$  0.7189  $  &  ~~~~~~~$ -0.5136 $  & ~~~~~ $ -0.8824 $     
\\
\hline 
\end{tabular}    
\end{center}
\end{table}
\begin{figure}\centering
    \subfloat[]{\includegraphics[scale=0.30]{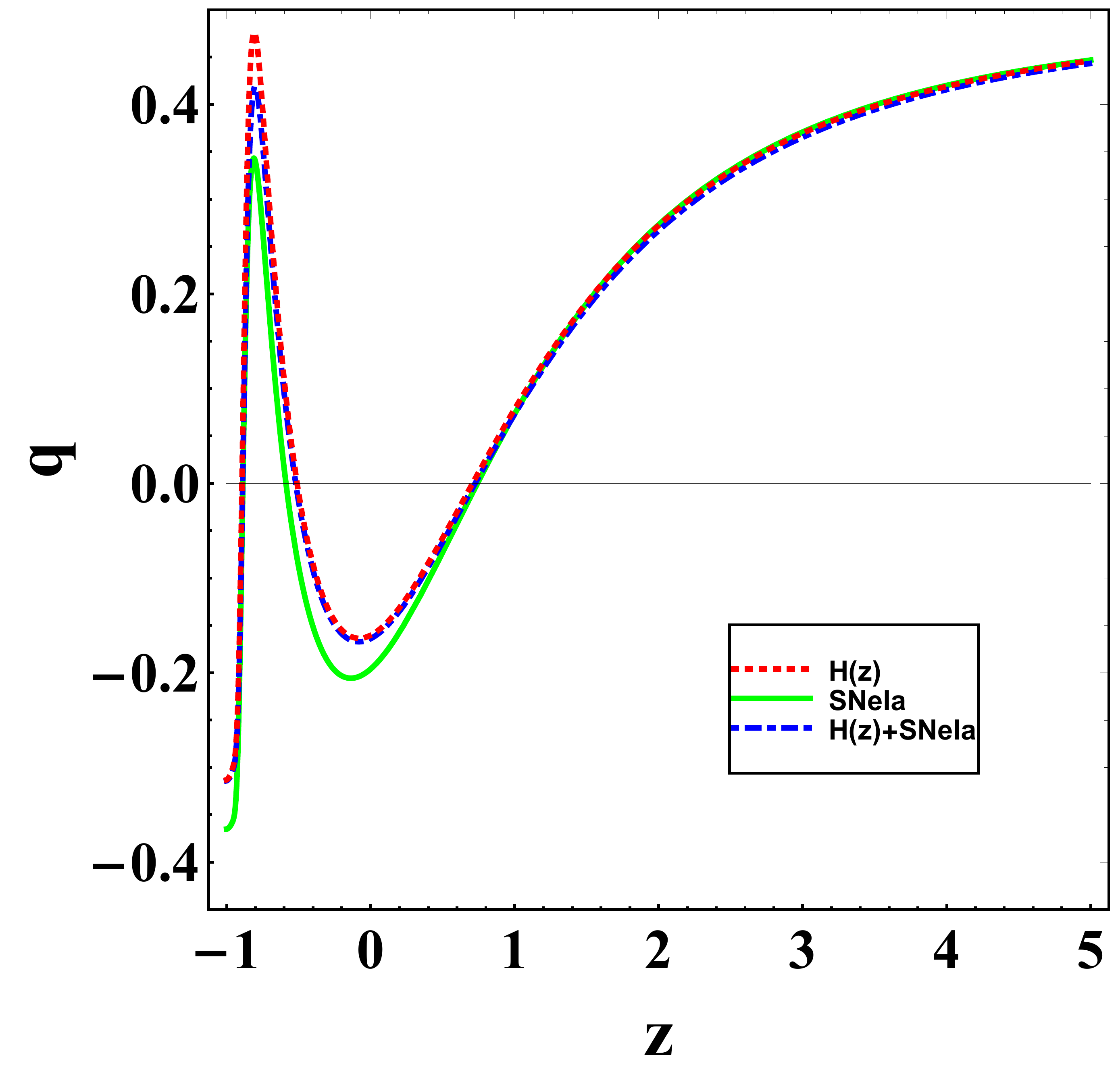}}\hfill
	\subfloat[]{\includegraphics[scale=0.30]{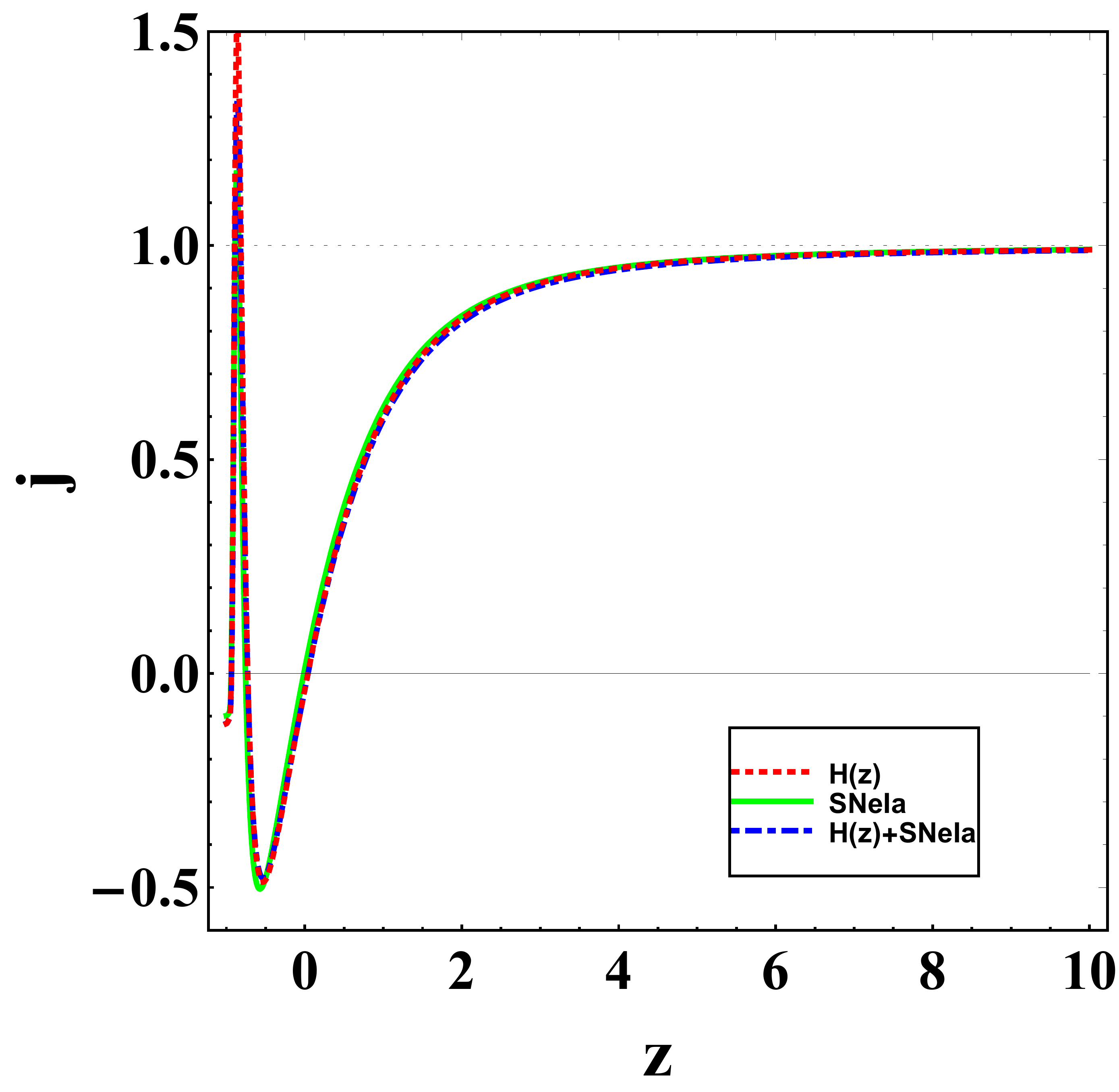}} 	
\caption{\scriptsize The plot of $ q $ and jerk parameter \textit{w.r.t.} $ z $.}
\end{figure}
To understand the evolution of the expanding universe in a more significant way, we explore some cosmographic parameters which contain the higher-order derivatives of scale factor $ a $. Here, we intend to discuss a kinematic quantity that contains the third-order derivative and is known as the jerk parameter. The Jerk parameter $ j $ is defined as
\begin{equation}\label{30}
j=\frac{\dddot{a}}{a H^3}=-q+2q(1+q)+(1+z)\frac{dq}{dz}.
\end{equation}
The standard value of the  Jerk parameter $ j $ for $ \Lambda $CDM model is one, thus digression from $ j=1 $ explores the evolution of different kinds of dark energy models. Fig. 6(b) highlights that at the early time our model is similar to $ \Lambda $CDM model and afterward this deviates from $ \Lambda $CDM till the late times.

To differentiate our model with $ \Lambda $CDM, we also use statefinder diagnostic technique, which is established by Sahni et al. \cite{Alam:2003sc, sah}. The two geometrical diagnostic parameters $ (s,r) $, known as statefinder parameters are defined as
\begin{equation}\label{31}
r=\frac{\dddot{a}}{aH^3} ~~~~~~ \and ~~~~~~ s=\frac{r-1}{3(q-\frac{1}{2})}, ~q\neq \frac{1}{2}
\end{equation}
Using statefinder diagnostic parameters, we can compare the goodness of various dark energy models with $ \Lambda $CDM. In $ s-r $ plot, the points $ (0,1) $ and $ (1,1) $ represent $ \Lambda $CDM model and $ SCDM $ respectively. From Eqs. (\ref{21}), (\ref{30}) and (\ref{31}), $ r $ and $ s $ can be calculated, and by using the best-fit values of model parameters we draw these trajectories (see Fig. 7(a)). The arrows on the curves show the direction of evolution. In our model $ r<1 $ and $ s>0 $ at an early time, which corresponds to the quintessence DE model. The present values of $ {(s,r)} $ are evaluated $ {(0.5214,-0.0330), (0.4729,-0.0125), (0.5190,-0.0338)} $ from $ OHD $, $ SNeIa $ and their joint datasets respectively, which show the deviations from $ \Lambda $CDM at present.

In Fig. 7(b), it is highlighted that the $ q-r $ curves start from $ SCDM $ $ (1,\frac{1}{2})$ and enter into the acceleration zone for all datasets in the future. In this plot, the dotted horizontal line $ r=1 $ denotes $ \Lambda $CDM. Here, we observe that the trajectories are crossing the transition line thrice from early evolution up to late times. As $ q $ is negative in late times, therefore this model shows an acceleration phase of the universe in the future.
\begin{figure}\centering
	\subfloat[]{\label{a}\includegraphics[scale=0.34]{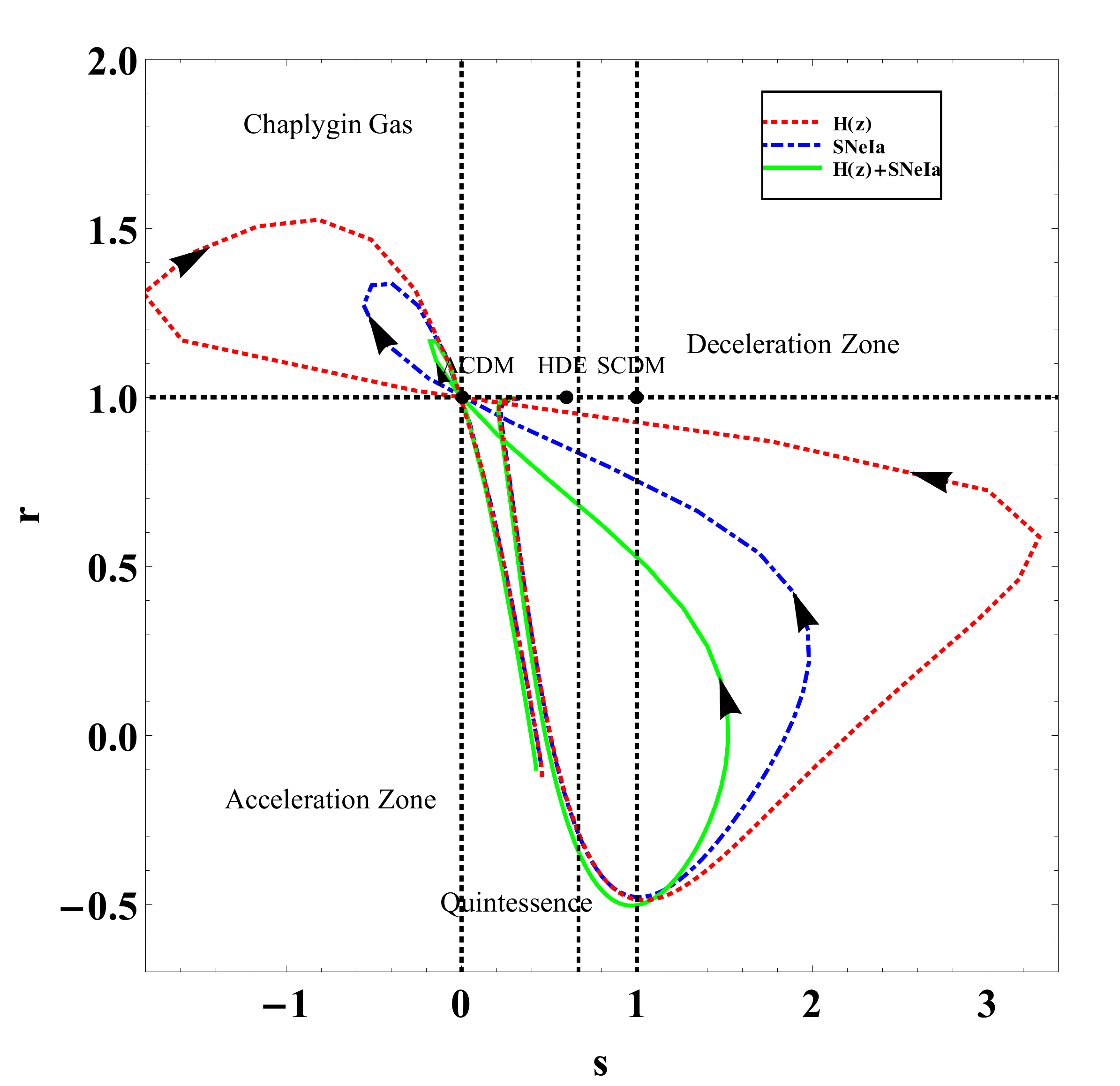}}\hfill
	\subfloat[]{\label{b}\includegraphics[scale=0.38]{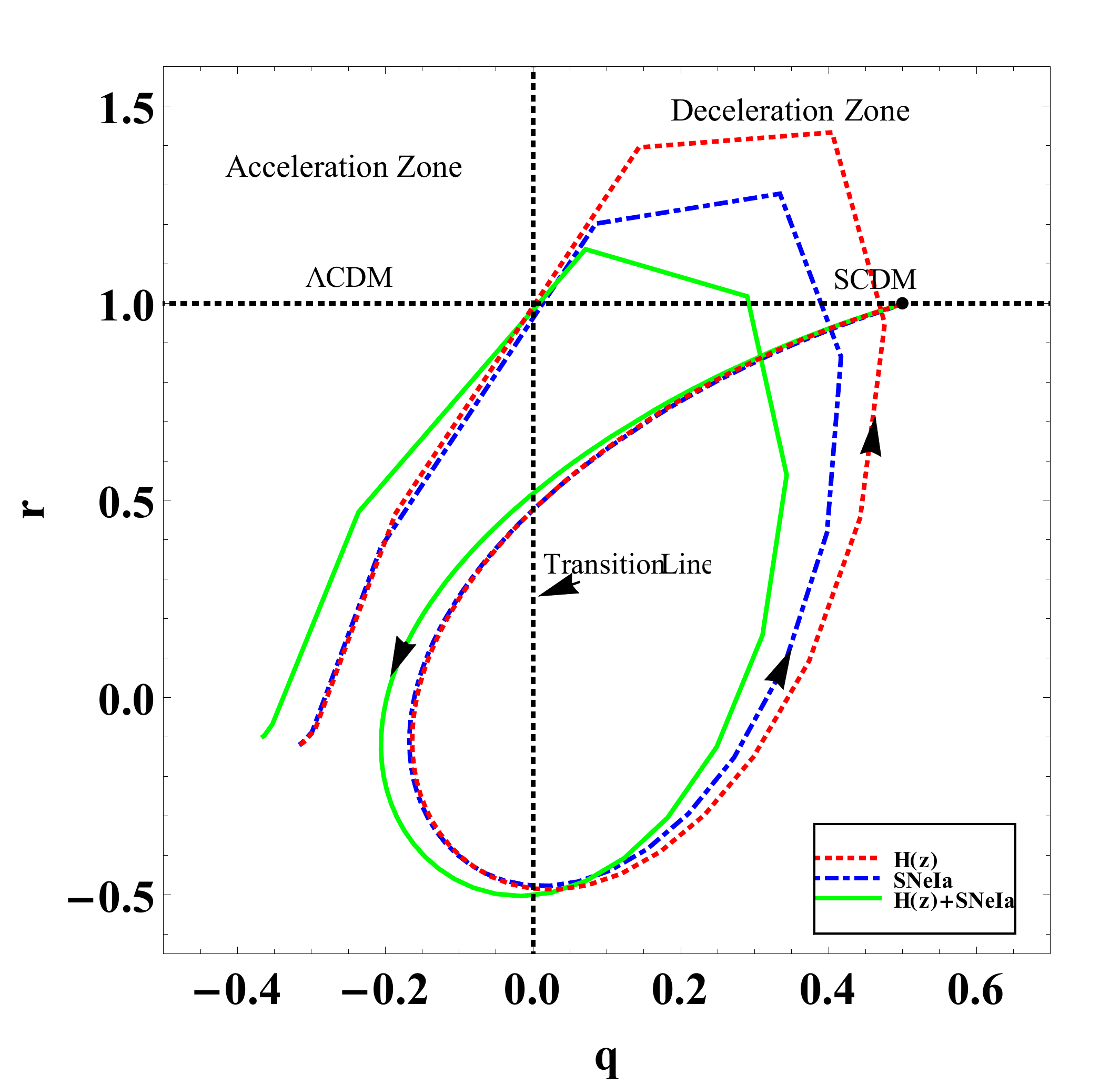}} 
	\caption{\scriptsize The Statefinder plots $ s-r $ and $ q-r $}
\end{figure}
\section{Conclusion}
\qquad In this model, we have explored the late-time cosmic expansion of the universe in the framework of a flat FLRW space-time metric with time-dependent displacement vectors based on Lyra’s geometry. We have derived EEF from the principal action (\ref{1}) and find solutions. Using the recent observational datasets $ OHD $, $ SNeIa $ and $ OHD+SNeIa $, we have obtained the best-fit values of model parameters $ \alpha $, $ \beta $ by applying MCMC method with \textit{emcee} package in Python. Furthermore, we investigated the behavior of cosmological parameters for the constrained values of model parameters, which are evaluated in Tables I, and II. 

The energy density of scalar field $ \rho_\phi $ decreases monotonically from high redshift to low redshift and $ \rho_\phi \rightarrow 0 $ as $ z\rightarrow -1 $, $ \forall $ observations. The pressure $ p_\phi $ is having negative finite value throughout the range, which indicates the accelerating behavior of the Universe (see Fig. 4). The EoS parameter of the scalar field $ \omega_\phi $ depicts that our model is a quintessence dark energy model as $ \omega_\phi $ is having a value between -1 to 0 for the large range of $ z $. Its value becomes positive for a while, and at the end, it converges to -1 for all datasets. In addition, it clearly highlighted the evolution profile of the deceleration parameter involving three transitions for each dataset which are shown in Table III. The jerk parameter shows that our model is similar to $ \Lambda $CDM in the early universe and it deviates in the future (see Fig. 5,6).
The statefinder diagnostic graphs show that for all datasets, $ s-r $ trajectories are looking in the form of loops in our model and passing through the $\Lambda $CDM model. Also, three $ q-r $ trajectories, which are also loops for the present model, starts from SCDM and show acceleration at its final phase (see Fig. 7). Thus, in this paper, we conclude that this model is an accelerated expanding model having the characteristics of the quintessence model studied in Lyra geometry, which is different from $ \Lambda $CDM model. It is relevant to point that the approach here presented could be applied in other theories of gravity, such as $f(R)$ \cite{Rinaldi:2014gua}, $f(R,T)$ \cite{Moraes:2016gpe}, $f(Q)$ \cite{BeltranJimenez:2017tkd,  Mandal:2020lyq}, and $f(Q,T)$ \cite{ Xu:2019sbp, Arora:2020met}. Such applications could give us extra parameters to test the viability of Lyra geometry with respect to experimental data. We hope to report these extra contributions in near future. 

\acknowledgments 
JRLS would like to thank CNPq (Grant no. 309494/2021-4), and PRONEX/CNPq/FAPESQ-PB (Grant nos. 165/2018, and 0015/2019) for financial support. JASF thanks FAPES for financial support.

\end{document}